\documentclass[amsmath,amssymb,aps,pre]{revtex4-1} 

\usepackage[american]{babel}
\usepackage[utf8]{inputenc}
\usepackage{subcaption}
\usepackage{graphicx}
\usepackage[hidelinks]{hyperref}
\usepackage{color}
\usepackage{floatrow}
\usepackage[table, dvipsnames]{xcolor}
\usepackage{booktabs}
\usepackage{array}

\begin{document}
\title{Trend analysis of the COVID-19 pandemic in China and the rest of the world}
\author{Albertine Weber}
\address{Instituto de Física, Universidade Federal do Rio Grande
	do Sul, Porto Alegre, RS, Brazil}
\affiliation{Cappra Institute for Data Science, Porto Alegre, RS, Brazil}

\author{Flavio Iannelli}
\affiliation{URPP Social Networks, University of Zürich, Andreasstrasse 15, CH-8050 Zürich, Switzerland}

\author{Sebastián Gonçalves}
 \email{sebastiangoncalves@gmail.com}
\affiliation{Instituto de Física, Universidade Federal do Rio Grande
	do Sul, Porto Alegre, RS, Brazil}
\affiliation{URPP Social Networks, University of Zürich, Andreasstrasse 15, CH-8050 Zürich, Switzerland}

\date{\today}
\begin{abstract}
The recent epidemic of Coronavirus (COVID-19) that started in China has already been ``exported" to more than 140 countries in all the continents, evolving in most of them by local spreading.
In this contribution we analyze the trends of the cases reported in all the Chinese provinces, as well as in some countries that, until March 15th, 2020, have more than 500 cases reported. Notably and differently from other epidemics, the provinces did not show an exponential phase.
The data available at the Johns Hopkins University site~\cite{JHU} seem to fit well an algebraic sub-exponential growing behavior as was pointed out recently~\cite{Meiner-2020}. All the provinces show a clear and consistent pattern of slowing down with growing exponent going nearly zero, so it can be said that the epidemic was contained in China.
On the other side, the more recent spread in countries like, Italy, Iran, and Spain show a clear exponential growth, as well as other European countries. Even more recently, US ---which was one of the first countries to have an individual infected outside China (Jan 21st, 2020)--- seems to follow the same path.
We calculate the exponential growth of the most affected countries, showing the evolution along time after the
first local case. 
We identify clearly different patterns in the analyzed data
and we give interpretations and possible explanations for them.
The analysis and conclusions of our study can help countries that, after importing some cases, are not yet in the local spreading phase, or have just started.
\end{abstract}
\maketitle

\section{Highlights}

\begin{itemize}
    \item All the provinces of China show very similar epidemic behaviour. 
    \item Early  stages  of spreading can  be  explained  in  terms  of  SIR  standard  model, considering that reported cases accounts for the removed individuals, with algebraic growing (sub-exponential) in most locations.
     \item Worldwide, we observe two classes of epidemic growth: sub-exponential during almost all stages (China and Japan) and exponential on the rest of the countries, following the early stage.
    \item The exponential growth rates ranges from 0.016$day^{-1}$ (South Korea) to 0.725$day^{-1}$ (Brunei)  which means 1.6\% to 107\% of new cases per day, for the different countries but China.
\end{itemize}

\section{Introduction}
The COVID-19 pandemic~\cite {WHO}, recorded daily by Johns Hopkins University (JHU)~\cite {JHU-database}, accounted for almost 180,000 cumulative cases worldwide. Approximately 81,000 of them have been reported in China, while Italy is the second country in the number of cases, with more than 27,000. Those numbers correspond to March 15th, 2020.

\begin{figure}[H]
	\centering
	\includegraphics[width=1\textwidth]{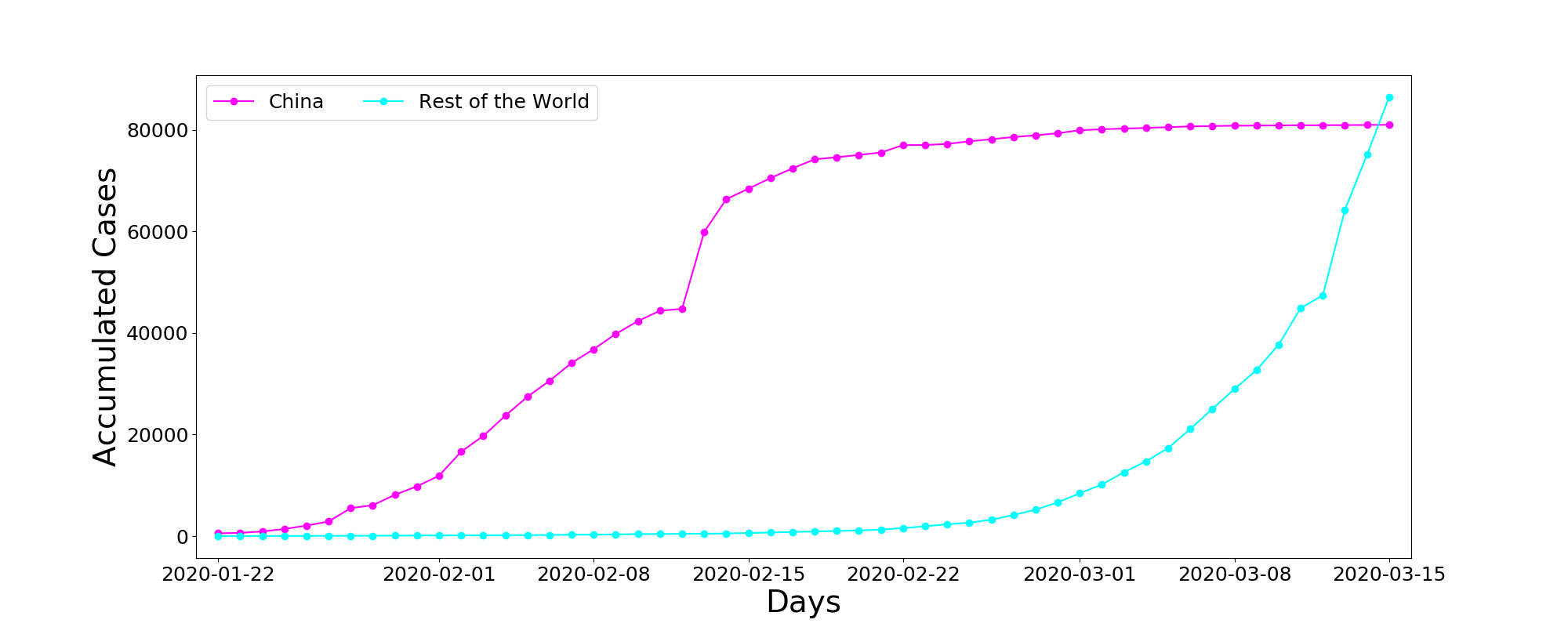}
	\caption{Accumulated Cases for China vs all the other countries, from first recorded day at JHU, Jan 22nd,  until  March 15th, 2020.}
	\label{fig:china_vs_world}
\end{figure}

The disease has already spread to more than 140 other countries ---and is likely to reach all countries--- thus, understanding its behavior is a priority. In Fig.~\ref{fig:china_vs_world} we see the evolution of the cases detected with the virus in China and the rest of the world, with clearly very different trends. As China has asymptotically reached 81,000 cases, the sum of cases from all other countries has exceeded that number on March 15th, 2020, after 55 days of the first case outside of China.

As local outbreaks are still in the early stages in most countries, identifying and analyzing the propagation patterns in countries with a considerable number of cases is crucial in helping others prepare, anticipate and, hopefully, avoid the overload point of each national health system.

\section{Methods}
The infection from the COVID-19 virus produces flu symptoms in most cases, with an asymptomatic but potentially infecting phase.~\cite{Rothe}. Regardless of that, the natural history of this disease can be modeled in principle as to SIR or SIRS compartmental models.
For the present times, in which the pandemic is just starting, and before there is evidence supporting the possibility of reinfection, we assume a SIR model to describe the dynamics of the local spreading for every country.
Then, at the early stages, it can be approximated by the linear differential equation:
\begin{align}
\frac{dI}{dt} = \beta SI - \gamma I\approx (\beta  - \gamma)  I ,
\label{sirlinear}
\end{align}
where $I$ and $S$ are the density of infected an susceptible, respectively. 
The solution of Eq.~\eqref{sirlinear} has the form $I(t) \approx I(0) e^{(\beta-\gamma)t}$.
In this framework, therefore, an exponential growing phase is expected with growth rate $r=\beta-\gamma$, from where the infectiousness parameter $\beta$ or the basic reproductive number $R_0$ could be derived, if we know the recovery rate $\gamma$. 

Such behavior of the number of infected happens only when there is no intervention at all. Also, it refers to the number of infected of a given epidemic, which is in general unknown.
However, in the ongoing pandemic of COVID-19, there is a continuous, daily account of the number of confirmed cases, worldwide.
And the identified individuals, after tested positive for the Corona virus, are then isolated or set to quarantine, {\em i.e.}, they are {\em removed} from the dynamics. They actually account for the $R$ compartment, which is governed instead by the differential equation $\frac{dR}{dt} = \gamma I$, with early-stage solution:
\begin{align}
R(t) \approx 
\frac{\gamma I(0)}{(\beta-\gamma)}
\left(e^{(\beta-\gamma)t}-1 \right)
\approx
at + bt^2
\label{removedtofit}
\end{align}
where $a=\gamma I(0)$ and $b=\gamma I(0)(\beta-\gamma)/2$.


Based on the above arguments, in the next Section, we analyse the time series of the 32 provinces/regions of China, assuming that the increase of cases is of algebraic type. However, we do not force the fit to the expression \eqref{removedtofit}, we fit with a $N=a t^b$ relation instead, and obtain the resulting exponents.
The fitting of the power-law is made in a moving window of seven days, and plot the $b$ exponent as a function of day for the whole time series, starting at day one for each province and ending on March 12th (as March 15th is the last day considered in this study and each value correspond to the center of the 7 days window). 

For the international outbreak, we first applied the parabolic fitting of Eq.~\eqref{removedtofit} to three countries (Germany, Italy, and Japan) and compare them with China (Fig.~\ref{fig:4countries}). After that, we dismiss the algebraic fitting and use the exponential growth ($I(t) \approx I(0) e^{(\beta-\mu)t} = A e^{Bt}$), from Eq.~\eqref{sirlinear}, instead. Consequently, we fit an exponential function in a moving window of seven days and plot the $B$ growth rate as a function of day for the whole time series.

\section{Results}\label{res}
\begin{figure}[H]
	\centering
	\includegraphics[width=0.8\textwidth]{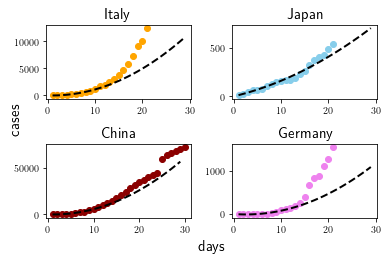}
	\caption{Reported cases during the first month after onset of spreading phase in four countries. From top left to bottom right, in Italy, Japan, China and Germany since Feb 21th or the first outbreak day (for China) and the corresponding  algebraic  fitting $R(t)\approx at+bt^2$ compatible with early-time expansion of the unconstrained SIR model, see Eq.~\eqref{removedtofit}}
	\label{fig:4countries}
\end{figure}

Before delving into more specific analysis, we first observe in Fig.~\ref{fig:4countries} the results of the algebraic fit $R(t)\approx at+bt^2$ in comparison with the real data for four different countries: China, Italy, Japan and Germany.

The patterns observed in Fig.~\ref{fig:4countries} suggest that different models should be applied according to the country that is being analysed.
Clearly, Italy and Germany deviate from the algebraic behavior and become exponential after about 15 days from the outbreak, signaling the substantial delay with respect to China and Japan of the implementation of containment strategies. For that reason, we divided the analysis: for China~\footnote{While the parabola, according to Eq.~\ref{removedtofit}, fit very well the early stages of China data, an exponential fit for a very shot period is valid too. However, in the long term, the exponential has to change its growth rate continuously as the algebraic is better for longer periods.} we focused on the sub-exponential model; for the other countries, we focused on the exponential fit.

\subsection{China Provinces/Regions}
\begin{figure}[H]
	\centering
	\includegraphics[width=1\textwidth]{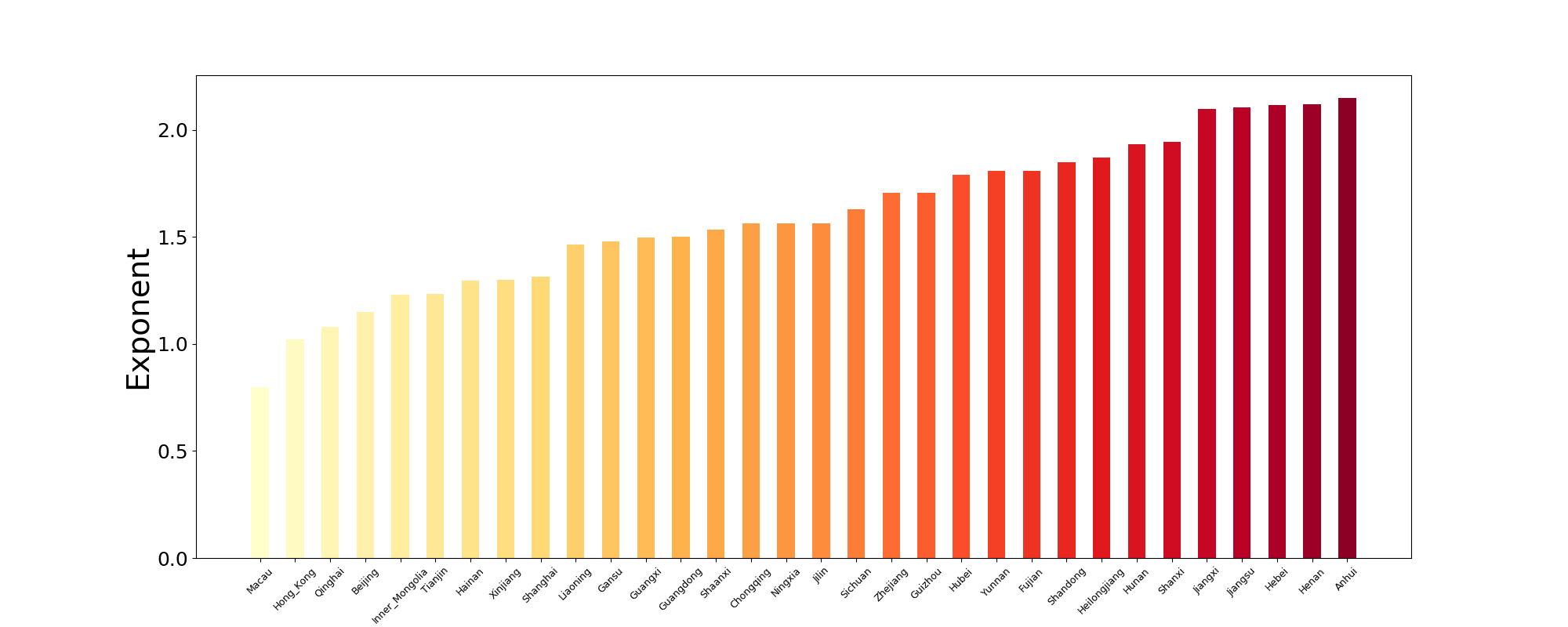}
	\caption{Coefficients of algebraic fitting, $n \sim t^b$, from day 1 to Feb 11th for the 32 provinces/regions of China.}
	\label{fig:coef-china}
\end{figure}

We start by presenting in Fig.~\ref{fig:coef-china} the coefficient $b$ of the fitted law $\text{Cases} = at^b$ for Chinese provinces. The fitting of the algebraic sub-exponential epidemic growth cover the period from Jan 22nd ---the first day of data--- to Feb 11th, 2020.

As we can see in Fig.~\ref{fig:coef-china}, the algebraic fitting provides consistent values in all provinces/regions, with exponent between 1 and 2, as stated by Eq.~\ref{removedtofit}. This is compatible with the hypothesis of the removing population  represented by the recorded number of cases.
Moreover, the fast isolation of infectives (turning them into removed, indeed), means increasing the rate $\gamma$, so effectively lowering the basic reproductive number $R_o$.

\begin{figure}[H]
	\centering
	\includegraphics[width=1\textwidth]{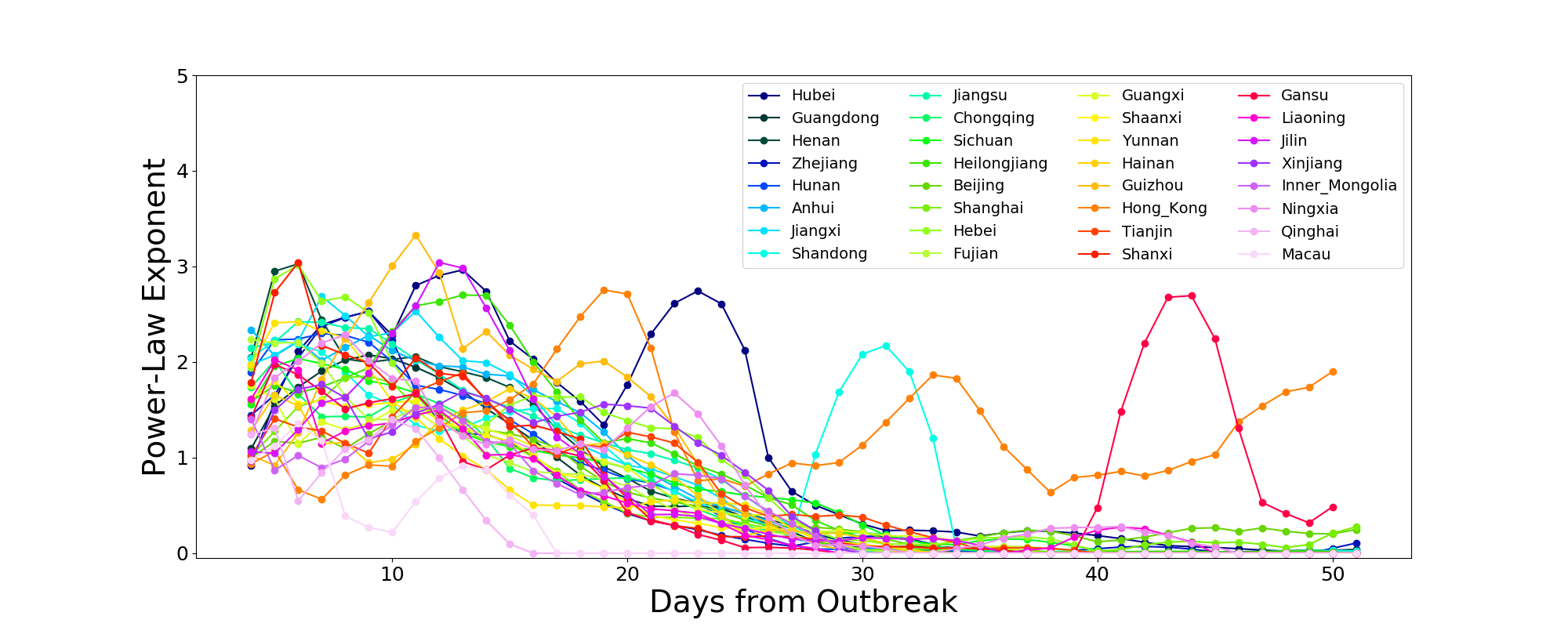}
	\caption{Coefficients of algebraic fitting, $n \sim t^b$, as a function of days after outbreak, based on 7 days moving window, for all the Chinese provinces. From Jan 22nd to March 15th, 2020.}
	\label{fig:power-china}
\end{figure}

Then, the power-law fitting is extended thorough a moving window of seven days, covering the period from Jan 22nd to March 15th, 2020. The results of the fits, for each of the province/regions of China, are shown in Fig.~\ref{fig:power-china}.

The following general characteristics are evident: there is a peak (some cases with exponent $b \approx 3$), mainly in the first two weeks after the local outbreak, then a clear and continuous decrease in the exponent of the algebraic fitting.
The slowing down to $b \approx 0$ is almost general, except for two locations: Gansu, with a late peak, and Hong Kong, where the epidemic seems to be bouncing up. Therefore, the epidemic is not fully contained even when total numbers (all China) do not show that so clearly. 
Those peaks at different times in different regions may indicate the epidemic arriving later or national measures implemented later. It is also important to stress that the peaks are of the exponents. 
As the growth rate for a power law regime $n \sim t^b$ is $\Delta n/n = b/t$,
an exponent 3 at day 30 gives a growth rate of 10\%, while an exponent 2 at day 10 represents
a growth rate of 20\%.
Regardless of that, the general trend of algebraic fitting, with most exponents going most zero, is a clear indication of a coordinated effort.

\subsection{International}
\begin{figure}[H]
	\centering
	\includegraphics[width=1\textwidth]{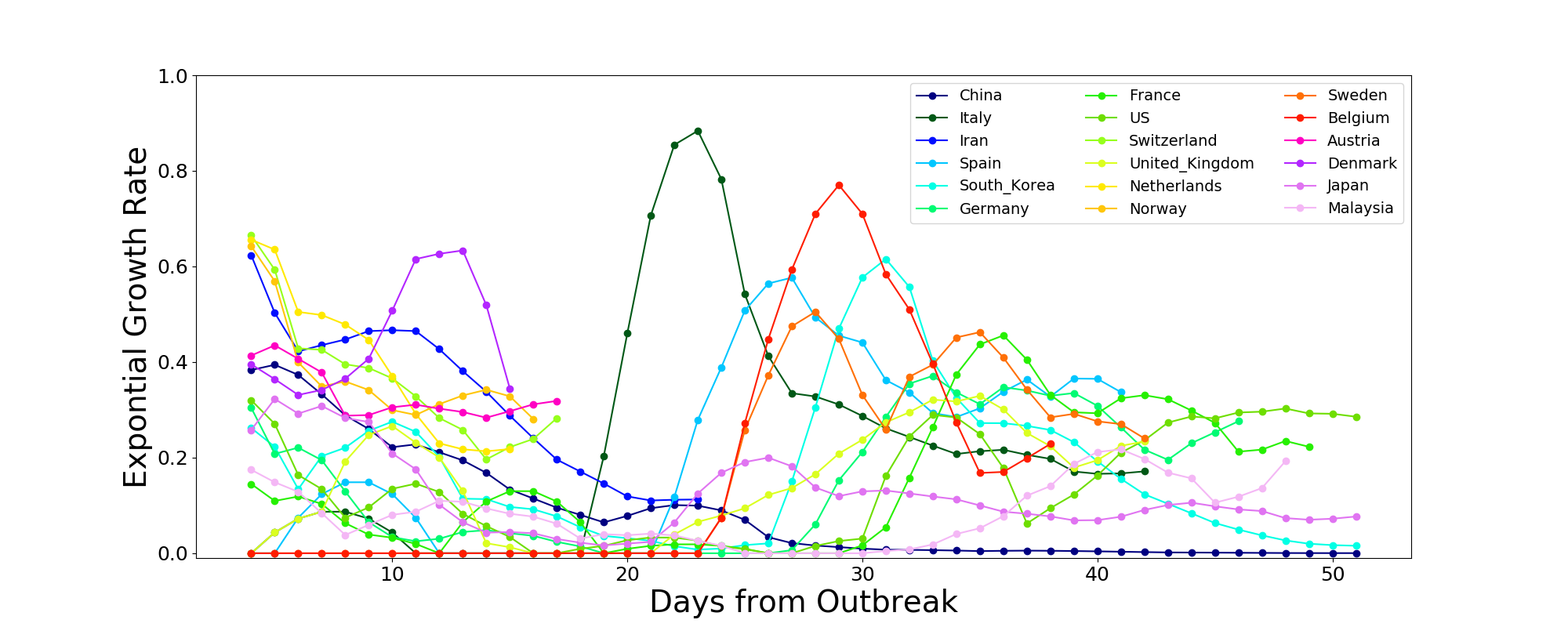}
	\caption{Growth rates from exponential fitting, as a function of days after local outbreak, evaluated on a moving window of 7 days, for the countries with more than 500 cases on March 15th, 2020.} 
	\label{fig:exp-inter}
\end{figure}

For the international outbreak, we make the seven day moving window analysis, but using the exponential growth $I(t) \approx I(0) e^{(\beta-\mu)t} = e^{Bt}$. We then plot in Fig.~\ref{fig:exp-inter} the $B$ rate as a function of day, for all the countries that have more than 500 cases on March 15th, 2020. Even when we have shown in the previous section that the data from China is well represented by an algebraic function, we include an exponential fitting of the same data for comparison with the other countries.

\begin{figure}[H]
\centering
  \begin{subfigure}{6cm}
    \includegraphics[width=5cm]{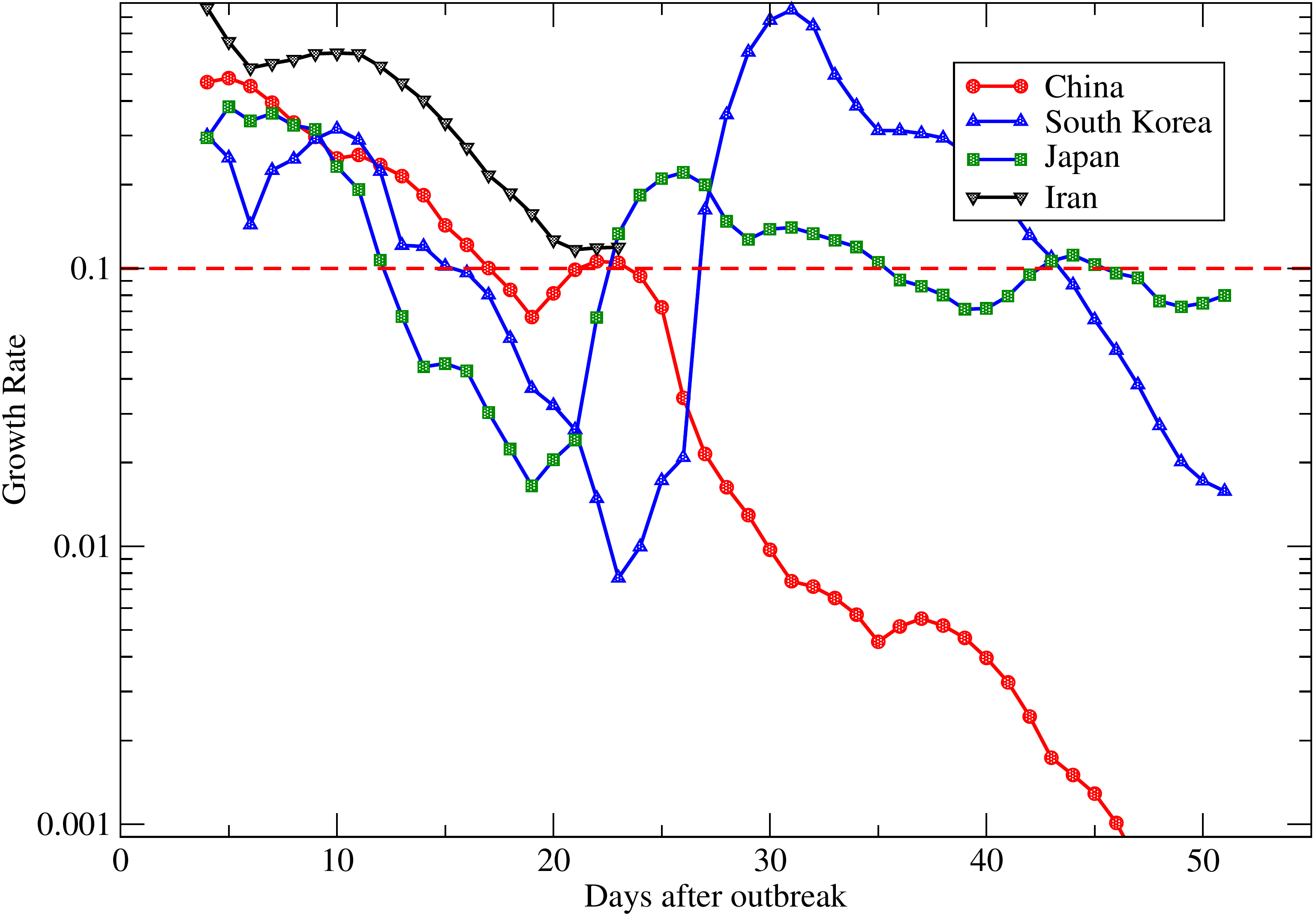}
    \caption{}
  \end{subfigure}
  \begin{subfigure}{6cm}
    \includegraphics[width=5cm]{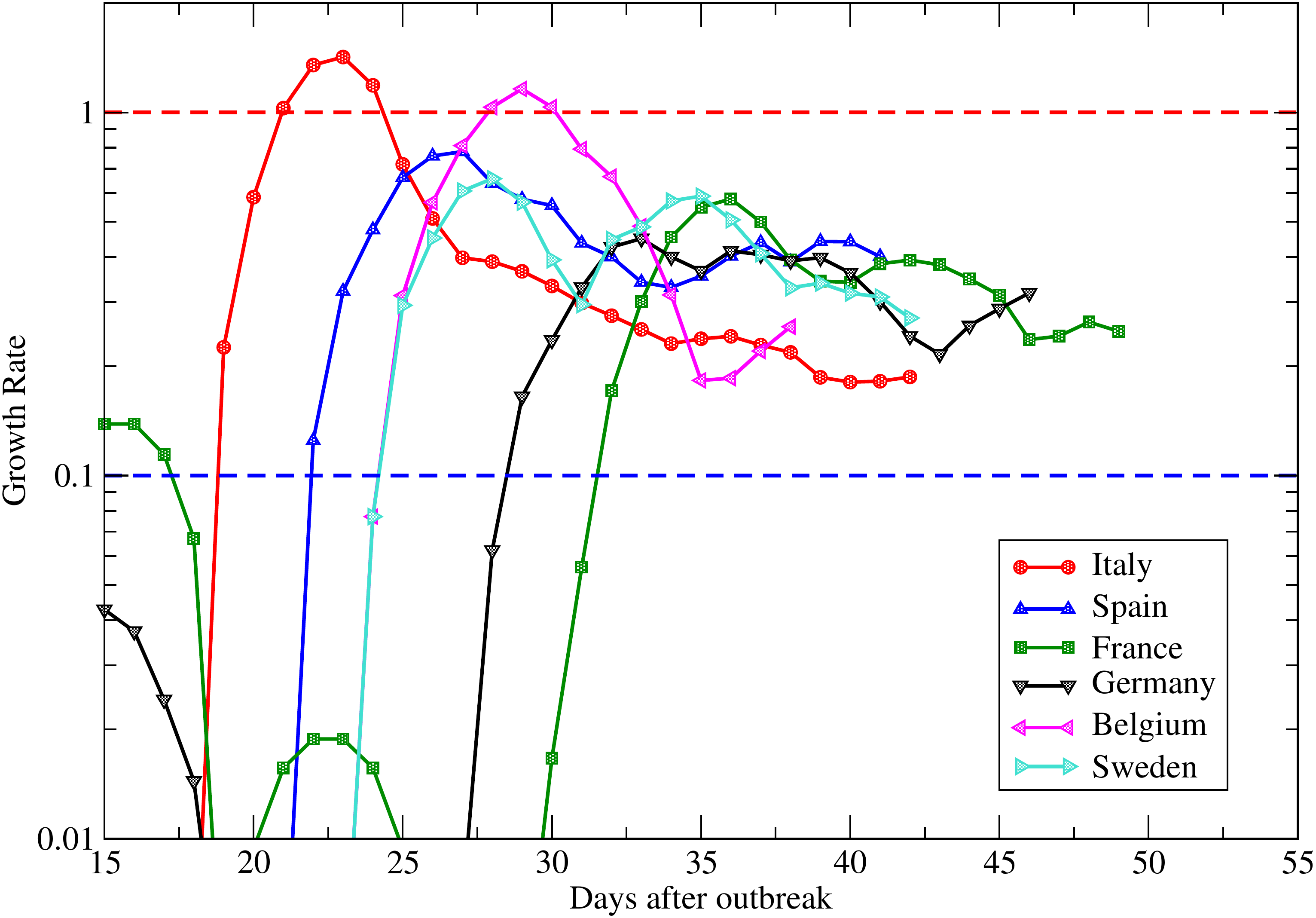}
    \caption{}
  \end{subfigure}
 
  \begin{subfigure}{6cm}
    \includegraphics[width=5cm]{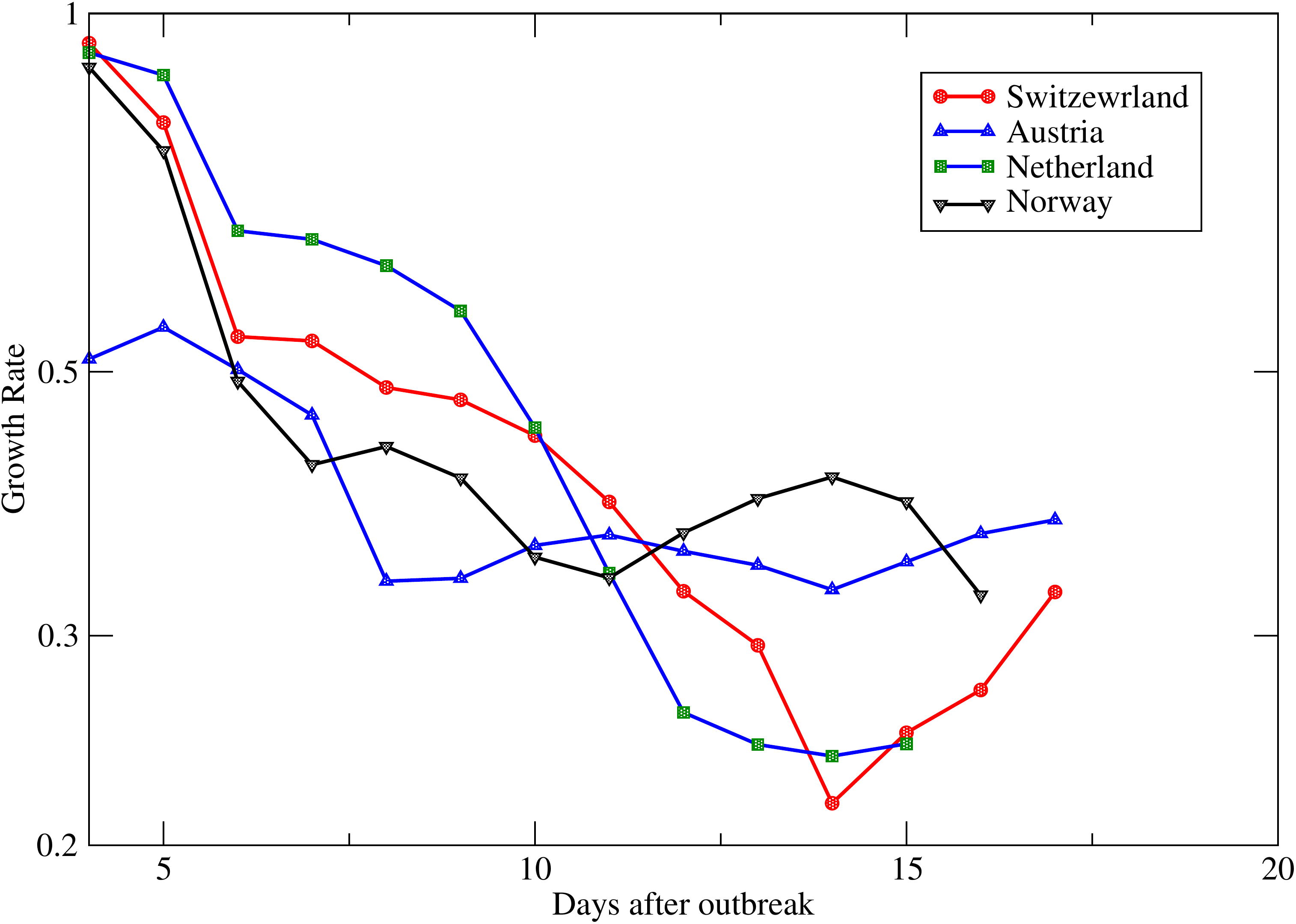}
    \caption{}
  \end{subfigure}
  \begin{subfigure}{6cm}
    \includegraphics[width=5cm]{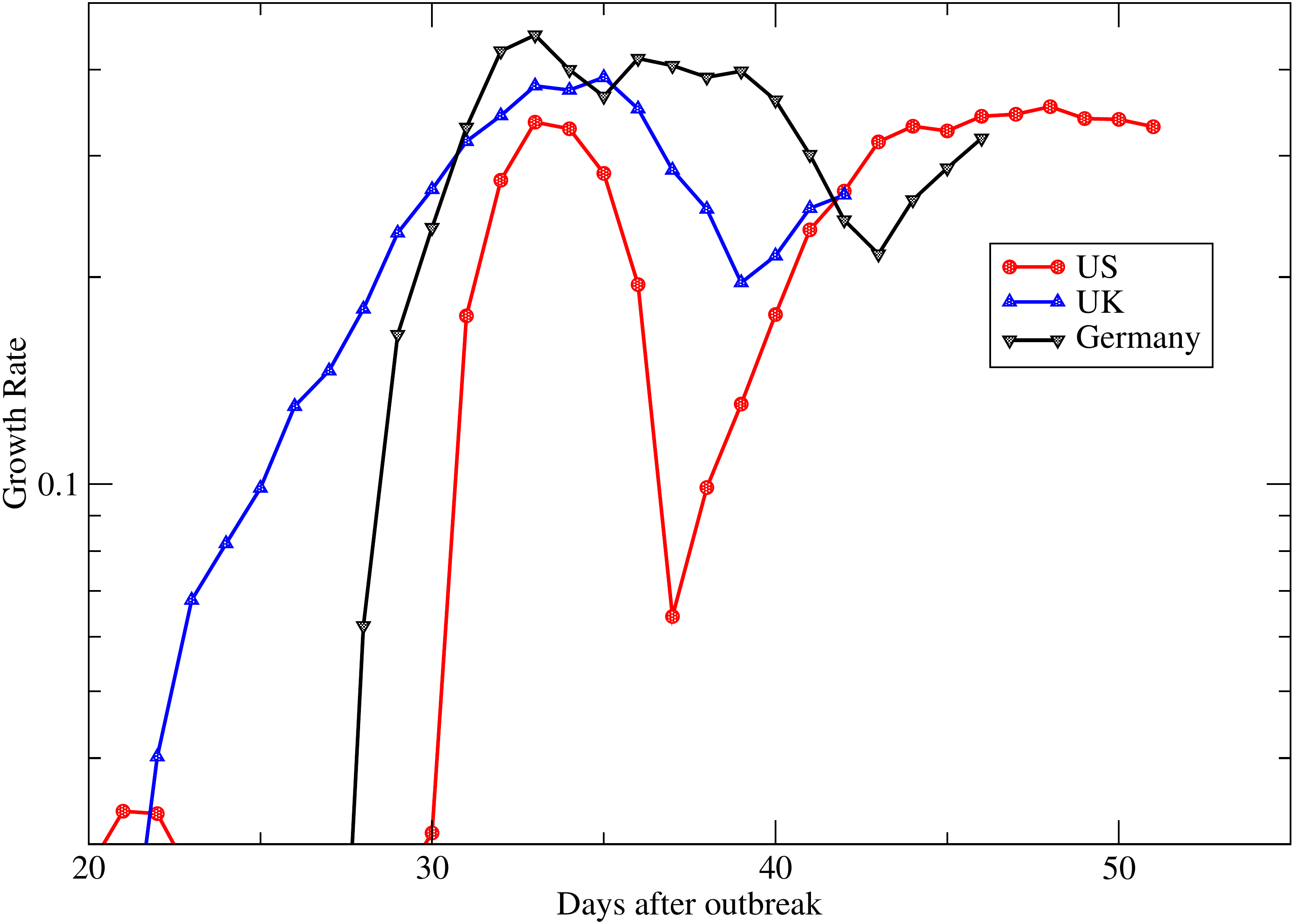}
    \caption{}
  \end{subfigure}
  \caption{Evolution of the Growth Rate per day for the most affected (more than 500 cases) countries by March 15th, 2020, From day one of each country until March 15th, 2020.
  Short scales (c) are due to countries with more recent outbreaks. (b) and (c) are displayed from day 15 and 20 respectively to highlight the more recent data.}
  \label{fig:4groups}
\end{figure}

The Fig.~\ref{fig:exp-inter} allows us to see the different behaviors of the COVID-19 spreading in those countries.  
Apart from some particular cases ---most distinctive China but South Korea, Japan, and maybe Iran too--- the rest seem to fit in one of two patterns, which we show in detail in Fig.~\ref{fig:4groups}. Either they have a dormant phase (mostly imported cases with no local infections), followed by the onset of the exponential local spreading inside the country (Fig.~\ref{fig:4groups}b), or they skip the dormant phase and enter the exponential phase almost immediately (Fig.~\ref{fig:4groups}c). The last behaviour occurs for many European countries that have delayed the arriving of infected cases and are still facing the first weeks of the outbreak. They are lowering the growth rate, apparently in a steady way, but the growth rate $\sim 0.3$ are still high, which means a doubling time of less than 2.5 days.

The four countries displayed in Fig.~\ref{fig:4groups}a have the most distinctive behavior from the rest.
The common denominator is the steady lowering of the growth rate coefficient. But the paths to the lowering are different between them. While for China is very clear, except for two small bumps in the third, and seventh week, due to  delayed burst in some regions (see Fig.~\ref{fig:power-china}), South Korea and Japan show a more erratic pattern, oscillating  around the $0.1$ until recently, when South Korea finally managed to plunge its rate, following now the Chinese pattern. Iran, in turn, has not reached yet the $0.1$ mark, but has been consistently lowering its rate since the beginning of cases, although it might be to soon to see if it is definitely going down. Finally, although Japan has managed to have less than 10\% daily rate, it seems to struggle yet to lower even more its rate.

Figure \ref{fig:4groups}d shows the data on the growth rate of the US and the UK. They belong to the same group as displayed in Fig.~\ref{fig:4groups}b, but we decided to separate for the sake of clarity and show together with Germany as an example of (b) group.  We see that they are at high values after a rebound in a qualitatively same pattern as Germany. The difference in the case of the UK and US is that they had more fluctuations (not seen in the panel) before the clear onset of the local spreading phase.

\begin{figure}[H]
	\centering
	\includegraphics[width=1\textwidth, scale=1.5]{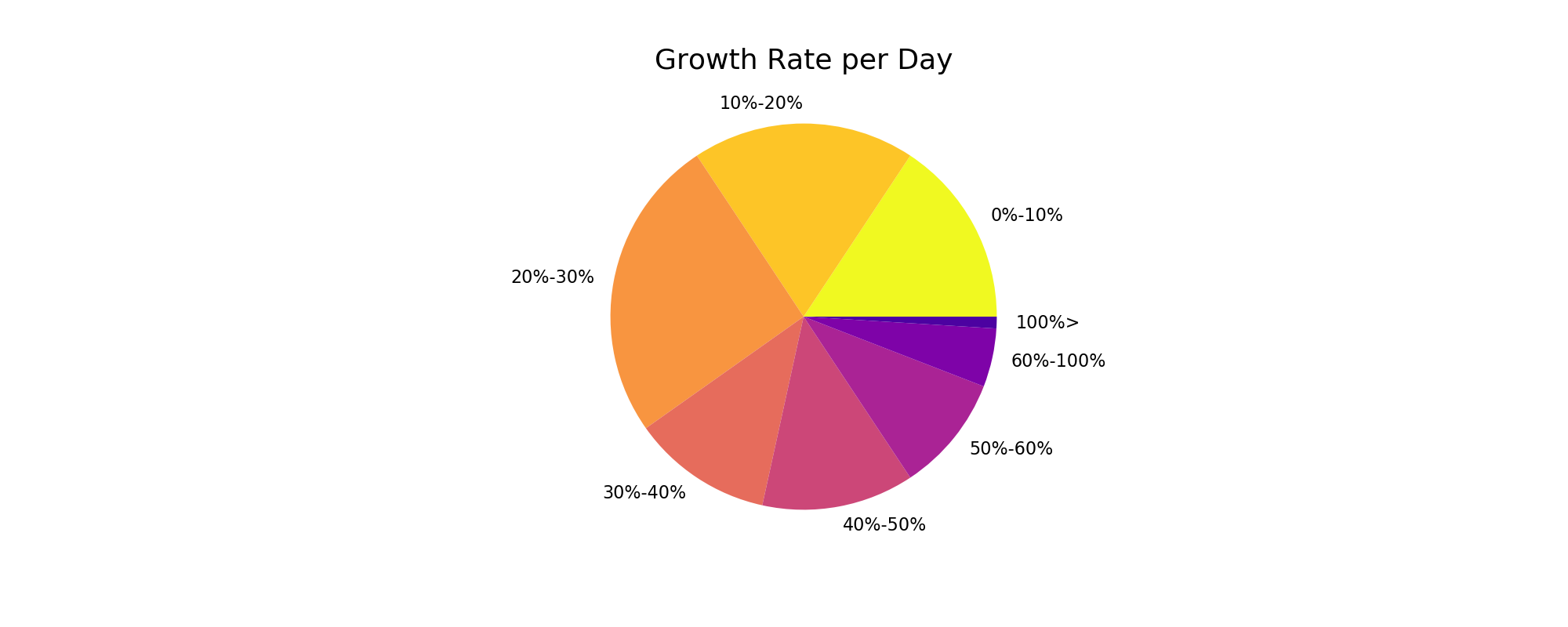}
	\caption{Pie-chart distribution of Growth Rate in percentage per  Day, for the week from March 9th-15th, 2020.
	The size of the pie-chart portions represent the fraction of countries inside each growth rate range.}
	\label{fig:growth-rate-groups}
\end{figure}

Applying the exponential fit to all the countries with at least seven days of confirmed cases, we can visualize the current situation of the COVID-19 outbreak. Fig.~\ref{fig:growth-rate-groups} shows the worldwide distribution of the Percentage Growth Rate per Day based in the week of March 9th-15th, 2020.
About some of the countries that have received more coverage in the news recently, either because were the first to show cases, or have a large numbers of cases, or high daily rates:
China is of course in the lower rate group, as it has virtually contained the spreading of the virus. Japan is also in this group, with 8\% spreading rate. In the second group there is Italy with 19\% rate, after having rates over 100\% more that 3 weeks after the arriving of the first case.
The group under 30\% is the more populated, with many EU countries like France (24\%), Sweden (28\%), and Germany (29\%). In the next group we have Switzerland (30\%) and US (33\%). In the group between 41\% and 50\%, we find Brazil (41\%), after 18 days from first case, and Finland (46\%), 46 days after its first case. In the sixth group, we have South Africa (54\%) and Denmark (55\%). Over 60\%, we have Colombia (70\%) and Qatar (72\%). Finally, Brunei (107\%) is the only country over 100\%.
For the complete list of countries, see Table~\ref{tab:rates} in the Appendix.

\section{Discussion}

The analyzed data of the pandemic of COVID-19 in the period from Jan 22nd - March 15th, 2020, points out two different types of epidemic growth: sub-exponential and exponential. The sub-exponential pattern is represented mainly by China with almost all its regions, South Korea after exiting from the exponential regime, and maybe Japan and Iran.

\begin{figure}[H]
	\centering
	\includegraphics[width=0.6\textwidth]{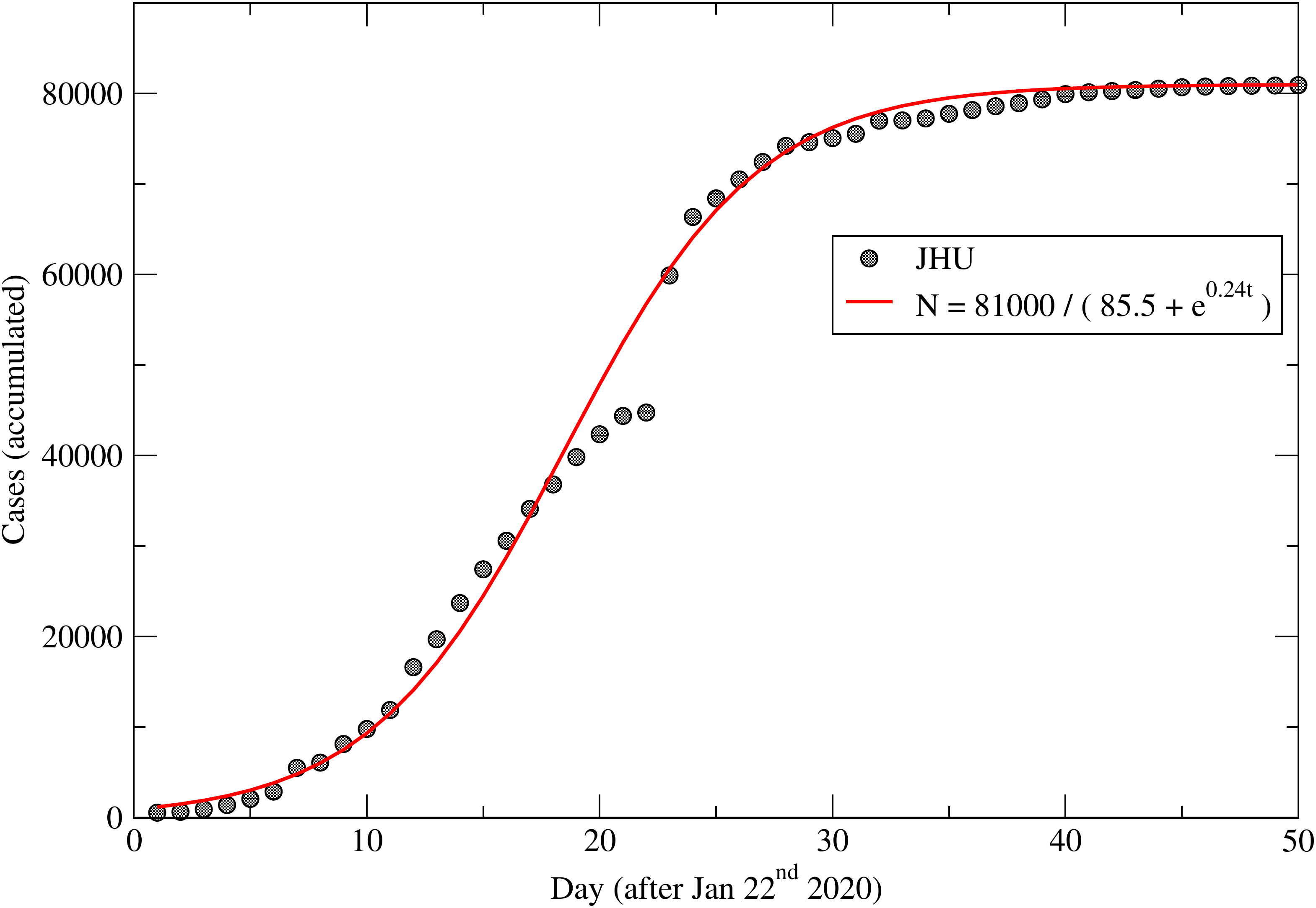}
	\caption{Accumulated number of cases in China (data from Johns Hopkins University \cite{JHU-database}, as a function of days after first reporting day (Jan 22nd, 2020) and fitting with a sigmoid function, correlation coefficient $R=0.96$.}
	\label{fig:china-data}
\end{figure}

The sub-exponential growing associated to a continuously slow of the exponent single out the result of simultaneous implementation of measures in China that made possible the relatively fast containment of the outbreak, stabilizing the number of cases. As a result, two months after the first official WHO report~\cite{WHO1st}, the total number of reported cases in China is excellently represented by a sigmoid, as we can see in Fig.~\ref{fig:china-data}.

Another important fact about China is that all its provinces/regions follow the same epidemic pattern. This is an evidence that China was able to contain the outbreak through a series of nationwide coordinated measures, corroborating the findings of WHO's joint mission in China \cite{WHO-joint-mission} and setting China as the most successful case in dealing with the COVID-19 epidemics so far.

Apart from these exceptions, the remaining countries appear to follow an exponential pattern. But even within the exponential framework we can observe different behaviours: either the countries have a dormant phase followed by an exponential stage or they already start on the exponential stage. In both cases, these patterns lead towards a quicker growth on the number of cases than those seem in China. Besides that, the scenario described by Fig.~\ref{fig:growth-rate-groups} is rather concerning: half of the countries are increasing cases counts on a 20\% to 50\% daily base, and almost one sixth is experiencing a growth even faster than this.

Some comments regarding the basic reproductive number $R_o$, and its estimation from 
the data here analyzed: the exponential growth rates from Fig.~\ref{fig:growth-rate-groups} are related with the parameters $\beta$ (infectiveness) and $\gamma$ (rate of removed) of the SIR model (sol. of Eq.~\eqref{sirlinear}) as  $B = \beta - \gamma$. As $\gamma=1/\tau$, we end up
with $B\tau = R_0 - 1$, where  $R_0 = \beta \tau$. Then $R_0 = 1 + B\tau$. 
Assuming that $\tau$ can be in the range $5$-$10$ days, we estimate
$R_0$ for three selected countries. With the lowest growth rate $B = 0.016day^{-1}$ for South Korea, we obtain $1.08 < R_0 < 1.16$, nearly to the epidemic threshold. On the other hand, for the biggest growth rate $B = 0.725day^{-1}$ for Brunei, we obtain $4.6 < R_0 < 8.3$. On the intermediate growth rates seeing in some EU countries, Sweden has  $2.2 < R_0 < 3.5$, in agreement early estimates of this number that put it around 3~\cite{Read}, and with other estimation~\cite{WHO46th}. However, if we take Italy at its worst day, we get $R_0$ as large as 10. 

Although worsening of the international scenario is expected, there are other countries out of China that can be look as examples of how to contain this epidemics. In fact, we observe that it is possible to slow the epidemic spread in both exponential frameworks: a few European countries - like Netherlands and Switzerland - started directly in the exponential stage and were still able to quickly reduce the magnitude of theirs coefficient. On the other hand, both South Korea and Italy had a spike after the ending of their dormant phase and have significantly lowered their growth rate per day after the adoption of containment measures like massive testing (South Korea) and lock-downs (Italy).
In the case of Italy the situation is much worse because the lower of the rate is not enough: still at 20\% rate and with more than 27,000 cases, if no further lowering is  attained quickly, the nation health system can soon be overwhelmed.

There is still a growing pandemics of COVID-19 to be contained. For that, the two months epidemic in China and the more recent outbreaks in Europe and Asia already serve as examples for the rest of nations to decide on how to manage this urgent and vital problem. The pattern observed in the current most affected countries reinforces the need to immediately implement massive, effective and coordinated containment strategies all around the world. Social distancing, isolation, quarantine have already mentioned as key-base~\cite{Anderson}.
The massive use of protection when interaction can not be avoided is another measure used in China and South Korea. And ultimately, temperature controls at house's members as was done in China to have a detailed account of the numbers of cases has to be considered too.

\section{Acknowledgments}
SG acknowledges CAPES for support, under fellowship  \#003/2019 - PROPG - PRINT/UFRGS, and URPP Social Networks and University of Zürich for hospitality.

\section{Apendix}
\begin{table}[H]
    \centering
    \begin{tabular}{c|l}
    Rate range (\%) & Countries \\
    \hline\hline
         \rowcolor{yellow!40} $\;0 - 10$
         & \multicolumn{1}{m{9cm}}{\tiny Andorra, Bangladesh, Bhutan, Cameroon, China, Holy See, Japan, Kuwait, Nepal, New Zealand, Nigeria, Oman, Singapore, South Korea, Taiwan, Togo}\\
         \hline
         \rowcolor{YellowOrange!40}$+10 - 20$  & \multicolumn{1}{m{9cm}}{\tiny  Algeria, Azerbaijan, Bahrain, Dominican Republic, Ecuador, Egypt, Georgia, India, Iran, Iraq, Italy, Jordan, Malaysia, Maldives, Monaco, San Marino, Thailand, United Arab Emirates, Vietnam}\\
         \hline
         \rowcolor{orange!40}$+20 - 30$  & \multicolumn{1}{m{9cm}}{\tiny  Afghanistan, Argentina, Australia, Belarus, Belgium, Cambodia, Canada, Costa Rica, Croatia, France, Germany, Greece, Hungary, Iceland, Israel, Lebanon, Liechtenstein, Martinique, Mexico, Netherlands, North Macedonia, Russia, Senegal, Sweden, Ukraine, United Kingdom}\\
         \hline
         \rowcolor{Red!40}$+30 - 40$  & \multicolumn{1}{m{9cm}}{\tiny  Austria, Bosnia and Herzegovina, Czechia, Ireland, Malta, Norway, Pakistan, Paraguay, Peru, Portugal, Switzerland, US}\\\hline
         \rowcolor{OrangeRed!40}$+40 - 50$  & \multicolumn{1}{m{9cm}}{\tiny  Brazil, Chile, Finland, Indonesia, Latvia, Lithuania, Morocco, Philippines, Poland, Romania, Saudi Arabia, Spain, Tunisia}\\
         \hline
         \rowcolor{RedViolet!40}$+50 - 60$  & \multicolumn{1}{m{9cm}}{\tiny Albania, Bulgaria, Cyprus, Denmark, Estonia, Moldova, Slovakia, Slovenia, South Africa, Sri Lanka}\\\hline
         \rowcolor{Plum!40}$+60 - 100$ & \multicolumn{1}{m{9cm}}{\tiny  Armenia, Colombia, Luxembourg, Qatar, Serbia}\\
         \hline
         \rowcolor{Blue!40}$+100 - $   & \multicolumn{1}{m{9cm}}{\tiny Brunei}\\
         \hline
   \end{tabular}
    \caption{Complete list of countries grouped according to the range of their spreading for the Covid-19. The values correspond to the week from March 9th - 15th, 2020, for countries with at least seven days from first reported case.}
    \label{tab:rates}
   \end{table}


\begin{thebibliography}{99}
\bibitem{JHU}
An interactive web-based dashboard to track COVID-19 in real time. The Lancet Infectious Diseases, S1473-3099, Vol 0, Issue 0 (2020). \href{https://doi.org/10.1016/S1473-3099(20)30120-1}{doi.org/10.1016/S1473-3099(20)30120-1}.

\bibitem
{Meiner-2020} Maier B, Brockmann D. Effective containment explains sub-exponential growth in confirmed cases of recent COVID-19 outbreak in Mainland China, (2020) \href{https://doi.org/10.1101/2020.02.18.20024414}{doi.org/10.1101/2020.02.18.20024414}.

\bibitem{WHO} 
Coronavirus disease 2019 (COVID-19) Situation Report – 51. March 11st, 2020. Available online: \href{https://www.who.int/docs/default-source/coronaviruse/situation-reports/20200311-sitrep-51-covid-19.pdf?sfvrsn=1ba62e57_10}{https://www.who.int/docs/default-source/coronaviruse/situation-reports/20200311-sitrep-51-covid-19.pdf}.

\bibitem{JHU-database}
JHU CSSE Team. 2019 Novel Coronavirus COVID-19 (2019-nCoV) Data Repository by Center for Systems Science and Engineering. JHU CSSE 2020.
Available online: \href{https://github.com/CSSEGISandData/COVID-19/blob/master/csse\_covid_19\_data/csse\_covid\_19\_time\_series/time\_series\_19-covid-Confirmed.csv}{https://github.com/CSSEGISandData/COVID-19/blob/master/csse\_covid\_19\_data/csse\_covid\_19\_time\_series/time\_series\_19-covid-Confirmed.csv} (accessed on 15 March 2020).

\bibitem{Rothe}
Rothe C et al., Transmission of 2019-nCoV Infection from an Asymptomatic Contact in Germany, N Engl J Med 382:970-971 (2020),
\href{https://doi.org/10.1056/NEJMc2001468}{doi.org/10.1056/NEJMc2001468}.

\bibitem{WHO1st}
Coronavirus disease 2019 (COVID-19) Situation Report – 1. January 21st, 2020. Available online: \href{https://www.who.int/docs/default-source/coronaviruse/situation-reports/20200121-sitrep-1-2019-ncov.pdf?sfvrsn=20a99c10_4}{https://www.who.int/docs/default-source/coronaviruse/situation-reports/20200121-sitrep-1-2019-ncov.pdf}.

\bibitem{WHO-joint-mission}
Report of the WHO-China Joint Mission.on Coronavirus Disease 2019 (COVID-19).
Available online: \href{https://www.who.int/docs/default-source/coronaviruse/who-china-joint-mission-on-covid-19-final-report.pdf}{https://www.who.int/docs/default-source/coronaviruse/who-china-joint-mission-on-covid-19-final-report.pdf}.

\bibitem{Read}
Read JM, Bridgen JRE, Cummings DAT, Ho A, Jewell CP. Novel coronavirus 2019-nCoV: early estimation of epidemiological parameters and epidemic predictions (2020). \href{https://doi.org/10.1101/2020.01.23.20018549}{doi.org/10.1101/2020.01.23.20018549}.

\bibitem{WHO46th}
Coronavirus disease 2019 (COVID-19) Situation Report – 46. March 06th, 2020. Available online: \href{https://www.who.int/docs/default-source/coronaviruse/situation-reports/20200306-sitrep-46-covid-19.pdf?sfvrsn=96b04adf_4}{https://www.who.int/docs/default-source/coronaviruse/situation-reports/20200306-sitrep-46-covid-19.pdf}.

\bibitem{Anderson}
Anderson RM, Heesterbeek H, Klinkenberg D, Hollingsworth TD.  How Will Country-based Mitigation Measures Influence the Course of the COVID-19 Epidemic? Lancet (2020). \href{https://doi.org/10.1016/S0140-6736(20)30567-5}{doi.org/10.1016/S0140-6736(20)30567-5}.

\end{thebibliography}
\end{document}